# Time-Delayed Reservoir Computing Based on Dual-Waveband Quantum-Dot Spin-Polarized Vertical Cavity Surface-Emitting Laser

M. Skontranis[1], G. Sarantoglou[1], A. Bogris[2], C. Mesaritakis[1]

[1]*University of the Aegean, Dept. of Information and Communication Systems Engineering, Palama 2, Karlovassi, 83200-Samos-Greece.*
[2]*University of West Attica, Dept. of Informatics and Computer Engineering. Aghiou Spiridonos, 12243, Egaleo, Athens, Greece.*
mskontranis@icsd.aegean.gr

**Abstract** In this work we present numerical results concerning a time-delayed reservoir computing scheme, where its single nonlinear node, is a Quantum-Dot spin polarized Vertical Cavity Surface-Emitting Laser (QD s-VCSEL). The proposed photonic neuromorphic scheme, exploits the complex temporal dynamics of multiple energy states present in quantum dot materials and uses emission from two discrete wavebands and two polarization states, so as to enhance computational efficiency. The benchmark task used for this architecture, is the equalization of a distorted 25Gbaud PAM-4 signal after 50Km of transmission at 1550nm. Results confirm that although typical ground-state emitting quantum-dot nodes offer limited performance, due to bandwidth limitations; by exploiting dual emission, we achieved a one-hundred-fold improvement in bit-error rate. This performance boost can pave the way for the infiltration of quantum-dot based devices in high-speed demanding neuromorphic driven applications.

## 1. Introduction

Reservoir Computing (RC) is a neuromorphic scheme that mimics structures present in the mammalian brain [1]. From an operational point of view, it originates from typical recursive neural networks (RNNs), but it is differentiated, by keeping its hidden layer node connections random and untrained. Through this step, RCs can significantly relax the complexity of the training procedure of RNNs and accommodate hardware-related imperfections without impacting computational performance [2]. This inherent hardware friendliness, has rendered RC a highly sought neuromorphic paradigm especially by photonic platforms. In particular, photonic RCs merge photonic driven merits, such as low power consumption and minimum processing latency with the RC's efficiency, so as to tackle demanding applications; among which are: machine vision [3], spoken vowel recognition [4], pseudo-chaotic timeseries prediction [5] and optical signal equalization [6], directly in the optical domain. Up to now, photonic implementations, can be split in two major categories: spatial RCs and time-delay RC (TDRC). In spatial RC, every neural node is a discrete photonic component, allowing ultra-fast operation but imposing restrictions when node up scaling is required [7]. On the other hand, TDRC approaches utilize a single nonlinear node, combined with an external optical feedback loop. In this scenario, physical nodes are replaced by virtual counterparts that exploit time multiplexing [7-9]. In this way, the physical node/connectivity restrictions of spatial RCs are alleviated. Following this approach, implementations include: electro-optical schemes based on Mach-Zehnder interferometers [10], whereas more numerous TDRC approaches rely on all-optical, laser nodes such as: typical DFB [9], Vertical Cavity Surface Emitting Lasers (VCSEL) [11], Spin Polarized VCSEL [12-14] and micro-ring lasers [15]. These implementations have provided enhanced performance in benchmark tasks such as Mackey-Glass and Santa-Fe series, whereas in real-world tasks e.g. optical channel equalization due to dispersion induced power fading, TDRCs have offered a power-efficient alternative to digital signal processing [7-16].

On the other hand, a fundamental drawback which derives from the use of virtual nodes is the tradeoff between the number of nodes and processing latency. This trade-off, directly impacts real-time processing applications, due to the fact that the input streams should be time-stretched so as to match the delay time of the optical feedback loop. This step is mandatory so as to feed the input to all the virtual nodes and update the state of the TDRC. In order to relax this inevitable impediment, recent implementations aimed to increase the number of virtual nodes, while keeping the feedback loop short. This has been implemented by exploiting multiple longitudinal modes in Fabry-Perot lasers [9] or different polarization states in quantum well VCSELS [12-14]. In these cases, emission of each mode (longitudinal mode/polarization state) could act as a quasi-independent RC, thus multiplying the number of virtual nodes. This solution cannot scale, due to the common gain mechanism that enforce a high level of correlation among modes. In addition, these approaches were based on typical quantum well (QW) structures. Following this lead, quantum dot (QD) based devices exhibit unique features, compared to QWs that are highly sought in neuromorphic schemes. Enhanced temperature insensitivity, originating from the three-dimensional confinement of carriers in the nanostructure, can alleviate the need for power-hungry active cooling schemes in QD based neuromorphic schemes [17]. Furthermore, recent works, have demonstrated that active QD layers can be grown directly over silicon, providing high-quality lasers [17], thus avoiding complex hybrid integration processes [18]. More importantly QD lasers, are known to exhibit emission from multiple wavebands, due to the easy access to stimulated emission from higher energy bands; namely ground-state (GS) and excited state (ES). In addition, the emission from these states is governed by complex inter band carrier dynamics, associated with different transition times [19], Paoli blocking effects [20], even different linewidth enhancement factor [21]. Therefore, these bands can be exploited as complex RC "modes" allowing a performance improvement compared to single band emitting devices.

In this work, we present numerical results concerning a spin polarized QD VCSEL, able to simultaneous emit from the GS and ES and from two polarization states. These four quasi-independent fields are utilized, for the first time according to our knowledge, in a TDRC scheme, allowing increased virtual node count. Through this approach, we circumvent the limited bandwidth of QD lasers and achieve enhanced performance, in many cases comparable with state-of-the-art QW multi-mode Fabry-Perot RCs, but with reduced hardware complexity and lower power consumption [7,9]. The benchmark test regarded in this work consists of optical signal equalization of a 25Gbaud pulse-amplitude modulated signal with four states (PAM-4), subject to power fading effects due to the interplay of chromatic dispersion and photodiode's non-linearity. This application is chosen firstly because it is a highly sought industrial application [22] and secondly because it allows direct comparison with other photonic TDRCs. For this application, performance was increased by 20dB compared to the typical QD VCSELs and it was similar to state-of-the-art QW based RCs. Therefore, this work paves the way for exploiting the inherent merits of QDs in TDRC neuromorphic paradigms, alleviating bandwidth restrictions and providing access to QD materials to a new wide pallet of machine-learning applications.

This paper is organized as follows: in Section 2 the model used to simulate the QD material and the s-VCSEL is described in detail, alongside the overall architecture of the proposed neural network. Followed by section 3, where we present numerical results regarding the performance of the RC versus critical parameters and the potential of the proposed scheme as neuromorphic paradigm is discussed. Finally, section 4 summarizes the findings of this study.

## 2. Quantum-Dot material and Time-delayed Reservoir Computing architecture

*2.1 QD material numerical model*

In this work, the numerical model used to simulate the behavior of the single section InAs/GaAs QD s-VCSEL is based on the electron-hole QD Spin Flip Model presented in [23-25]. In this model, the conduction band of the QD s-VCSEL consists of three non-degenerate energy levels, designated as the Ground State (GS), the First Excited State ($ES_1$) and the Second Excited State ($ES_2$). These states are accompanied by an extra 2D Wetting Layer (WL) and are separated by 60meV. On the other hand, the valance band comprises five non-degenerate energy levels, one GS and four ES ($ES_1$, $ES_2$, $ES_3$ and $ES_4$) contrary to [23] and [24] which uses eight states, one GS and seven ES. This modification is done to simplify our model and it is in accordance with other electron-hole QD models such as [26]. The difference of the utilized model compared to other QD s-VCSEL models [27-28], is that it assumes non-degenerate states, similar to [23-25]. This simplification can affect the bias needed to trigger stimulated emission from each state, thus the only difference in terms of dynamics is the injection current level that should be applied so as to trigger multi-band operation and harness the same performance. Furthermore, the five energy states of the valance band are also accompanied with a 2D WL as it is done in the electron case while all the energy states of holes are separated by 10meV. As a result, the time evolution of electrons and holes is described by a set of ten rate equations, four for the electrons and six for the holes. However, in our analysis we have considered the spin orientation of the carriers (spin-up and spin-down) which doubles the number of the rate equations. Consequently, a set of twenty differential equations is deployed in order to fully describe the dynamics of the spin-oriented carriers in the QD spin VCSEL (eq.1-6). Without loss of generality, we assumed that the only permissible transitions of the carriers take place between adjacent states. The equations for the occupation probability of each state are:

$$\frac{dp_{GS}^{e,h\pm}}{dt} = \frac{(1-p_{GS}^{e,h\pm})p_{ES_1}^{e,h\pm}}{\tau_{ES_1 \to GS}^{e,h}} - \frac{(1-p_{ES_1}^{e,h\pm})p_{GS}^{e,h\pm}}{\tau_{GS \to ES_1}^{e,h}} - \frac{p_{GS}^{e\pm}p_{GS}^{h\pm}}{\tau_{sp}} \mp \frac{p_{GS}^{e,h+}-p_{GS}^{e,h-}}{\tau_{spin}^{e,h}}$$
$$- \frac{n_l v_g g_{GS}}{N_q}(p_{GS}^{e\pm} + p_{GS}^{h\pm} - 1)|E_{GS}^{\pm}|^2 \qquad (1)$$

$$\frac{dp_{ES_1}^{e,h\pm}}{dt} = \frac{(1-p_{ES_1}^{e,h\pm})p_{GS}^{e,h\pm}}{\tau_{GS \to ES_1}^{e,h}} - \frac{(1-p_{GS}^{e,h\pm})p_{ES_1}^{e,h\pm}}{\tau_{ES_1 \to GS}^{e,h}} + \frac{(1-p_{ES_1}^{e,h\pm})p_{ES_2}^{e,h\pm}}{\tau_{ES_2 \to ES_1}^{e,h}} - \frac{(1-p_{ES_2}^{e,h\pm})p_{ES_1}^{e,h\pm}}{\tau_{ES_1 \to ES_2}^{e,h}}$$
$$- \frac{p_{ES_1}^{e\pm}p_{ES_1}^{h\pm}}{\tau_{sp}} \mp \frac{p_{ES_1}^{e,h+}-p_{ES_1}^{e,h-}}{\tau_{spin}^{e,h}} - \frac{n_l v_g g_{ES_1}}{N_q}(p_{ES_1}^{e\pm} + p_{ES_1}^{h\pm} - 1)|E_{ES_1}^{\pm}|^2 \qquad (2)$$

$$\frac{dp_{ES_2}^{e,h\pm}}{dt} = \frac{(1-p_{ES_2}^{e,h\pm})p_{ES_1}^{e,h\pm}}{\tau_{ES_1 \to ES_2}^{e,h}} - \frac{(1-p_{ES_1}^{e,h\pm})p_{ES_2}^{e,h\pm}}{\tau_{ES_2 \to ES_1}^{e,h}} + \frac{(1-p_{ES_2}^{e,h\pm})p_{WL}^{e,h\pm}}{\tau_{WL \to ES_2}^{e,h}} - \frac{(1-p_{WL}^{e,h\pm})p_{ES_2}^{e,h\pm}}{\tau_{ES_2 \to WL}^{e,h}}$$
$$- \frac{p_{ES_2}^{e\pm}p_{ES_2}^{h\pm}}{\tau_{sp}} \mp \frac{p_{ES_2}^{e,h+}-p_{ES_2}^{e,h-}}{\tau_{spin}^{e,h}} - \frac{n_l v_g g_{ES_2}}{N_q}(p_{ES_2}^{e\pm} + p_{ES_2}^{h\pm} - 1)|E_{ES_2}^{\pm}|^2 \qquad (3)$$

$$\frac{dp_{ES_i}^{h\pm}}{dt} = \frac{(1-p_{ES_i}^{h\pm})p_{ES_{i-1}}^{h\pm}}{\tau_{ES_{i-1} \to ES_i}^{h}} - \frac{(1-p_{ES_{i-1}}^{h\pm})p_{ES_i}^{h\pm}}{\tau_{ES_i \to ES_{i-1}}^{e,h}} + \frac{(1-p_{ES_i}^{h\pm})p_{i+1}^{h\pm}}{\tau_{ES_{i+1} \to ES_i}^{e,h}} - - \frac{(1-p_{ES_{i+1}}^{h\pm})p_{ES_i}^{h\pm}}{\tau_{ES_i \to ES_{i+1}}^{h}}$$

$$-\frac{p_{ES_i}^{e\pm} p_{ES_i}^{h\pm}}{\tau_{sp}} \mp \frac{p_{ES_i}^{h+} - p_{ES_i}^{h-}}{\tau_{spin}^{h}} \quad (4)$$

$$\frac{dp_{WL}^{e\pm}}{dt} = \frac{I^{\pm}}{qAN_{WL}} + \frac{(1-p_{WL}^{e\pm})p_{ES_2}^{e\pm}}{\tau_{ES_2 \to WL}^{e}} - \frac{(1-p_{ES_2}^{e\pm})p_{WL}^{e\pm}}{\tau_{WL \to ES_2}^{e}} - \frac{p_{WL}^{e\pm} p_{WL}^{h\pm}}{\tau_{sp}} \mp \frac{p_{WL}^{e+} - p_{WL}^{e-}}{\tau_{spin}^{e,h}} \quad (5)$$

$$\frac{dp_{WL}^{h\pm}}{dt} = \frac{I^{\pm}}{qAN_{WL}} + \frac{(1-p_{WL}^{h\pm})p_{ES_4}^{h\pm}}{\tau_{ES_4 \to WL}^{h}} - \frac{(1-p_{ES_4}^{h\pm})p_{WL}^{h\pm}}{\tau_{WL \to ES_4}^{h}} - \frac{p_{WL}^{e\pm} p_{WL}^{h\pm}}{\tau_{sp}} \mp \frac{p_{WL}^{h+} - p_{WL}^{h-}}{\tau_{spin}^{h}} \quad (6)$$

where $p_X^{e\pm}$ and $p_X^{h\pm}$ denotes the electron and hole occupation probability of the $X^{th}$ energy state (X=GS, ES$_1$, ES$_2$, ES$_3$, ES$_4$, WL). The + and - stand for spin-down and spin-up orientation of carriers respectively while $\tau_{X \to Y}^{e(h)}$ is the time constant for the electron (hole) transition between energy level $X_1$ and $X_2$ where $X_{1,2}$=GS, ES$_1$, ES$_2$, WL with $X_1 \neq X_2$. $E_X^{\pm}$ is the complex optical field at the $X^{th}$ state where $E_X^+ (E_X^-)$ is associated with the recombination of spin-down (up) carriers. $E^+$ and $E^-$ are commonly referred as right and left polarized fields respectively [29]. The remaining parameters are included in Table 1 and their values are taken from [23-25,29] based on previously published works, which in turn are extracted by experimental investigation of such devices. From a physical point of view equations 1-3 are identical for electron and holes. In detail, the terms $\frac{p_i(1-p_{i-1,i+1})}{\tau_{i \to i-1,i+1}}$ of the equation are accountable for the transition of the carriers between two adjacent states. The terms $\frac{p_e p_h}{\tau_{sp}}$ and $\frac{v_g g_{ES_2}}{N_q}(p_{ES_2}^{e\pm} + p_{ES_2}^{h\pm} - 1)|E_X^{\pm}|^2$ simulates loss of carriers due to spontaneous and stimulated emission respectively while the terms $\frac{p^{\pm}-p^{\mp}}{\tau_{spin}}$ are responsible for the interaction of carriers between the two polarization states (spin-up and spin-down). Equation 4 simulates the behavior of the holes at ES$_3$ and ES$_4$ levels which are consider as non lasing states. This is the reason why the term of the stimulated emission is omitted. Finally, equations 5 and 6 simulate the exchange of carriers between the WL and ES$_2$ for the case of electrons and the WL and ES$_4$ for the case of holes.

Table 1. Parameters of the QD spin polarized VCSEL

| Symbol | Meaning | Value |
|---|---|---|
| $\tau_{WL \to ES_2}^{e}$ | Capture time from WL to ES$_2$ for electron | 2ps |
| $\tau_{ES_2 \to ES_1}^{e}$ | Capture time from ES$_2$ to ES$_1$ for electron | 2ps |
| $\tau_{ES_1 \to GS}^{e}$ | Capture time from ES$_1$ to GS for electron | 8ps |
| $\tau_{i \to i+1}^{e}$ | Electron relaxation time from lower to higher energy state | $\exp\left(\frac{\Delta E_{i \to i+1}^{e}}{k_B T}\right) \tau_{i+1 \to i}^{e}$ |
| $\tau_{i \to i-1}^{h}$ | Capture time from higher to lower energy state | 0.4ps |
| $\tau_{i \to i+1}^{e}$ | Hole relaxation time from lower to higher energy state | $\exp\left(\frac{\Delta E_{i \to i+1}^{h}}{k_B T}\right) \tau_{i+1 \to i}^{e}$ |
| $\tau_{rt}$ | Round trip time of the cavity | 0.12ps |
| $\tau_{sp}$ | Spontaneous radiative lifetime | 0.4ns |
| $\tau_{spin}^{e,h}$ | Spin relaxation for electron and holes | 400ps |
| $k_B$ | Boltzamann constant | 8.617×10$^{-5}$eV/K |
| T | Temperature | 300K |
| $u_g$ | group velocity | 8.45×10$^{7}$m/s |
| q | Electron charge | 1.6×10$^{-19}$C |
| $n_l$ | Number of layers | 7 |
| g | Gain of the GS, $ES_1$ and $ES_2$ | 11m$^{-1}$, 20m$^{-1}$, 10m$^{-1}$ |
| $N_{dot}$ | Quantum dot density | $5\times10^{10}$cm$^{-2}$ |
| $\Delta E$ | Distance between electron (hole) energy levels | 60meV (10meV) |
| $\rho$ | Radius of surface | 11μm |
| $a_Y$ | Losses at Y state | 500m$^{-1}$ |
| $\alpha$ | Linewidth Enhancement Factor | 1.5 |
| $\gamma_p$ | Birefringence rate | 15GHz |
| R1, R2 | Reflectivity of the facets of the QD spin VCSEL | 0.9988, 0.9944 |
| $T_{ext}$ | Length of the external loop | 240ps |
| $k_{inj}$ | Injection strength | 0.8 |
| $k_f$ | Feedback strength | 0.01 |

With regards to the output of the QD s-VCSEL, the photon density equation of the original model was substituted by an equivalent electrical field rate equation which was based on the Slow Varying Envelop Approximation (SVEA), similar to [29-30]. The purpose of this amendment is to include the detuning parameter which will be introduced later in this section. Consider that only GS, $ES_1$, $ES_2$ level support lasing operation as well as the spin orientation of the carriers, a total of six complex differential equations are required. These equations have the following form:

$$\frac{dE_Y^\pm}{dt} = \frac{1}{2}v_g\big(g_Y(p_Y^{e\pm} + p_Y^{h\pm} - 1) - a_Y\big)(1 + ja_{LEF_Y})E_Y^\pm - j\gamma_p E_Y^\mp \quad (7)$$

where $\alpha_Y$ is the internal losses of the Y$^{th}$ state with Y=GS, $ES_1$, $ES_2$, $\alpha_{LEFY}$ is the linewidth enhancement factor of the Y$^{th}$ state and $\gamma_p$ is the birefringence rate. The first term of equation 7 is responsible for the stimulated emission of the Y$^{th}$ state and the second term $j\gamma_p E_Y^\mp$ describes the interaction between the two polarized fields of the same states $E_Y^+$ and $E_Y^-$. In this work, the effect of spontaneous emission is omitted, which is a common practice in other models for QD spin polarization VCSELs [23-25] due to the fact that the main noise effect, originates from the thermal noise of the photodiodes at the output layer.

Equation 1-7 describe the operation of a free running laser under no external injection or feedback. However, a TDRC processor is based on a nonlinear node which is subject to external injection and self-feeding in order to process the incoming data. Consequently, to implement the QD s-VCSEL as a TDRC we must modify equation 7 so as to include an injection and a feedback term.

$$\frac{dE_Y^\pm}{dt} = \frac{1}{2}v_g\big(g_Y(p_Y^{e\pm} + p_Y^{h\pm} - 1) - a_Y\big)(1 + ja_{LEF_Y})E_Y^\pm(t) - j\gamma_p E_Y^\mp(t)$$
$$+ \frac{k_{injX}^\pm}{\tau_{rt}}E_{MX}^\pm(t)\exp(j2\pi df t) + \frac{k_{fX}^\pm}{\tau_{rt}}E_Y^\pm(t - T_{ext})\exp(j2\pi f_X^\pm T_{ext}) \quad (8)$$

Equation 8 can simulate the operation of the QD spin VSEL under injection and feedback. In detail the term $\frac{k_{injX}^\pm}{\tau_{rt}}E_{MX}^\pm(t)\exp(j2\pi df t)$ simulates the external injection to the QD spin VCSEL while $\frac{k_{fX}^\pm}{\tau_{rt}}E_Y^\pm(t - T_{ext})\exp(j2\pi f_X^\pm T_{ext})$ stands for the feedback. $E_{MX}^\pm$ denotes the injected optical field, $df$ is the frequency difference between the injected field and the internal field of the QD spin VCSEL, $\tau_{rt}$ is the round trip time of the cavity, $T_{ext}$ is the length of the external loop while $k_{inj}^{\pm}$ and $k_f^{\pm}$ are the injection and feedback coefficients respectively which are given by the following formula as in [31]:

$$k_{inj,f} = \left(\frac{1}{R} - R\right) \cdot r_{inj,f}$$

where R is the reflectivity of the facet of the VCSEL and $r_{inj}$ ($r_f$) is the ratio of the injected (feedback) field to the free running output of the QD spin VCSEL. The values of all the parameters used in this work are provided in Table 1.

Equations (1) to (8) are part of a complex model consisted of 26 rate equations. The complexity of our model originates from the inclusion of dedicated rate equations for holes, contrary to other published works [27-29] which simulated the behavior of QD spin VCSELs through the excitonic approximation. Although the latter formulation achieves good agreement with experimental works in conventional QD lasers [32-33], it has been shown that the inclusion of both electron and hole dynamics offer a more accurate description of experimental data (GS-ES transition/switching) [26,30,34], whereas at the same time offer interesting intraband dynamics (excitability, spiking etc.) [30]. Since accurate description of the transient QD dynamics are imperative towards the successful simulation of the Time Delay Reservoir Computing scheme we chose the more complex electron-hole Spin Flip Model proposed in [23-25] so as to fully describe the dynamics of QD spin VCSELs.

## 2.2 Neuromorphic Architecture

Conventional TDRCs, instead of using discrete components as neuromorphic nodes, employ a single non-linear node, usually an injection locked laser. This scheme's computational power, relies on the generation of multiple time-multiplexed virtual nodes using an optical feedback loop [7]. In particular, the virtual nodes are temporally distributed in a time interval *T* equally, whereas *T* and consequently the number of nodes (*n*) is regulated by the time length of the loop. This neuromorphic arrangement, although is hardware friendly it imposes an inevitable speed-penalty which reduces the processing speed of the TDRC by a factor $SP=T/T_s$, where $T_s$ is the duration of each sample at the input [7,8]. Therefore, so as to boost the computational power of the system a longer *T* is dictated but at the cost of SP. In addition, in order to decorrelate the response of each virtual node, a pseudo-random mask is applied at the input stream. In particular, each incoming symbol is sampled with a 100GHz photodiode and is represented with a four component vector d={$d_1,d_2,d_3,d_4$}. Following that, an oversampling of 3 is applied. Consequently, the mask m consists of 4×3=12 random values and it is implemented to every symbol at the input. Henceforth, the input to the TDRC is not the originally retrieved signal $d$, but a time-stretched version of it multiplied by a mask $m$ so as to match time-scale T. The spacing of the virtual nodes is set to $T_n$=20ps similar to [9] and the length of the external loop is set to $T_{ext}$=240ps. A schematic representation of the conventional TDRCs is depicted in fig.1a.

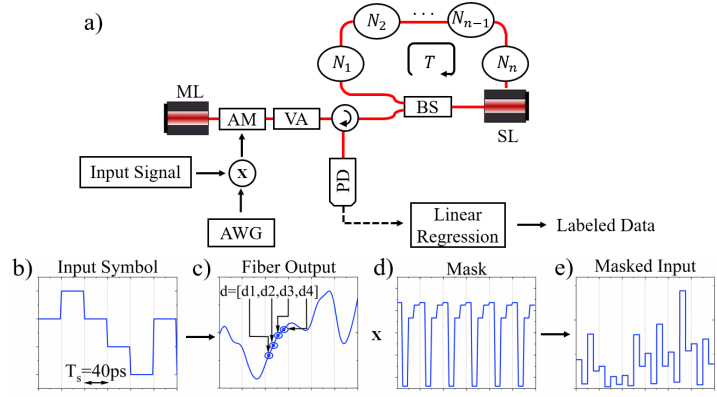

Fig. 1 a) Classical Time Delay Reservoir Computing (TDRC) with an external loop and a non linear node implemented in a Master Slave configuration. AM stands for amplitude modulator that allows the injection of the electrical input pattern to the scheme; VOA stands for variable optical attenuator that regulates injection strength. $N_i$ are the virtual nodes in the optical feedback loop, BS is a beam-splitter or fiber coupler and PD stands for photodiode. b) 25 Gbaud Pulse-amplitude-modulated signal with four levels (PAM-4). c) Retrieved 25Gbaud PAM-4 signal after a transmission of 50km. The retrieved signal is sampled with four samples per symbol d={$d_1,d_2,d_3,d_4$}. (d) 12 value mask repeated for every input symbol. (e) The input signal of the TDRC which is the dot product of the output of the fiber and the mask.

In this work, we aim to circumvent the *SP/n* tradeoff by exploiting the QD material properties. The proposed architecture is similar to previously studied TDRCs [7] but with some striking exceptions. Firstly, we aggressively reduce the length of the feedback loop to 20mm, assuming integrated silicon waveguide that corresponds to *T=240ps*, thus reducing *SP* and at the same time affect the number of nodes. Secondly, we replace convectional QW lasers with QD s-VCSELs in a Master-Slave laser (ML/SL) configuration. This step allows to explore the simultaneous use of the emission from the GS/ES states and from two polarization states/waveband so as to allow virtual-node increase beyond the feedback-loop imposed limit. The proposed scheme is depicted in Fig. 2. In detail, the continues wave (CW) emission of the ML is fed to a wavelength demultiplexer (DEMUX), which separates the GS/ES fields. After the DEMUX, two phase modulators (PM), are used so as to transfer *d·m* (masked input) from an arbitrary waveform generator (AWG) to the optical domain. The modulated signal from all bands is injected at the SL. The injection strength of each band is independently controlled by the two Variable Attenuators (VA). The SL, as in TDRCs, is subjected to feedback with the use of a short external loop. Finally, the output of the SL is guided to a set of DEMUX and polarization splitters (PSs) which drive four Photodiodes (PD). Each PD monitors a specific field of the SL ( $GS^+$, $ES^+$, $GS^-$ and $ES^-$). In order to render our simulations realistic performance wise we have included all typical noise effects at the output layer; including shot and thermal noise at the PDs. The detected photocurrent is sampled with 12 sample per symbol and quantized through a typical analogue-to-digital converter with 8bit resolution. Finally, the digital outputs are sent to an offline linear regression scheme.

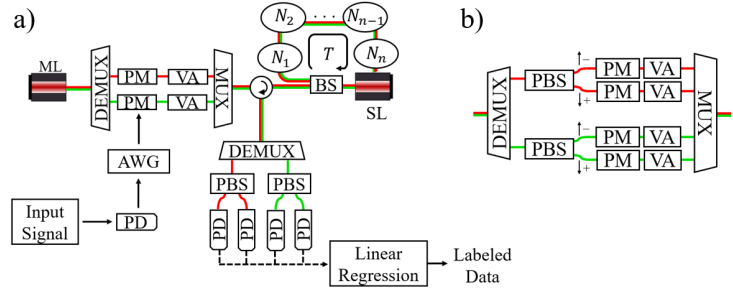

Fig. 2 a) Single-node time delayed reservoir computing scheme implemented by two QD spin polarized VCSEL. ML and SL are the Master and Slave Laser respectively, PBS is the polarization beam splitter, PM is a phase modulator, VA is a Variable attenuator, BS is a beam splitter accountable for the feedback, PD is a Photodiode and AWG is an Arbitrary Waveform Generator. b) Implementation with four different masks at the input. Each mask is applied to one polarized field of the dual emitting QD spin VCSEL. The arrows denote the Right Polarized field (+) related to spin down carriers and the Left Polarized Field linked (-) to spin up carriers.

## 3. Numerical Results

### 3.1 Solitary QD laser Emission

The aforementioned numerical model, was used in order to investigate typical characteristics of the solitary QD sVCSEL. Firstly, the relation between the applied bias current and emitted optical power was acquired by setting the injection and feedback coefficients in (8) equal to zero $(k_{inj} = k_f = 0)$. The results are presented in Fig. 3a. For simplicity, we plotted the sum of the output power of the left and right polarized field for each energy level and we assumed a common bias for both carrier orientations ($I^+ = I^-$). As it can be seen, the QD sVCSEL exhibits two lasing thresholds, one for the GS at $I_{GS,TH}=1mA$ and one for the $ES_1$ at $I_{ES1,TH}=4mA$, whereas stimulated emission from the second $ES_2$ was not observed for the bias range used. For this reason, for the rest of this work, the stimulated emission from $ES_1$ state will be referred simply as ES. The operational regime of the QD spin polarized VCSEL can be divided in three separate regions labeled as region I, region II and region III (Fig. 3a). In region I $(I < I_{GS,TH})$ no stimulated emission is observed. In region II ( $I_{GS,TH} < I < I_{ES,TH}$ ) the QD spin polarized VCSEL starts lasing from the GS state while ES emission is absent, while by increasing injection current within the limits of region II we noticed a constant increase at the GS output power. For bias higher than $I_{ES,TH} = 4mA$ the QD spin VCSEL is driven to region III, whereas ES state starts lasing while the GS level, which is almost filled with carriers ($p_{GS} \approx 0.9$), is saturated due to the Pauli Blocking effect. Moreover, further increase of the bias leads to increased ES power while the GS output power remains almost stable similar to [26,34].

The change in the slope of the GS emission at Fig.3 can be attributed to two main reasons: the increase of the GS occupation probability ($p_{GS}$) and the onset of ES emission [26,34]. First, the increase of the bias current augments $p_{GS}$. As $p_{GS}$ increases, the incoming rate of electrons towards the GS state decreases because of the term $1 - p_{GS}$ at the first term of equation (1). This term indirectly simulates the Pauli Blocking effect because when $p_{GS}=1$ the incoming rate of electrons becomes zero. Second, when ES starts lasing, most of the pumped carriers are consumed by the ES emission. This fact reduces the scattering rate from ES to GS even more. However, even though most of the pumped carriers are absorbed by the ES, there is still a small portion of them which ends up at GS state and augments $p_{GS}$ at equilibrium. These two facts are responsible for the slight increase in the GS emission when ES starts lasing. This behavior is also confirmed by other works [26,34].

The different operational regimes of the QD s-VCSEL are of particular interest, because when neuromorphic applications are considered, they can be associated with different temporal dynamics. In this context, it is established, that emission from different states is linked to different inter-band time constants [20,26,30], different linewidth enhancement factor [21], while under dual-band emission temporal correlation of the ES/GS can be adjusted [27,30]. On the other hand, QD devices are not massively adopted when high-speed operation is needed, due to their limited modulation bandwidth, which originates from the gain saturation and intra band carrier relaxation processes [35]. In order to shed light, we computed the 3dB modulation bandwidth of each state for regime III. It can be seen, that although the GS level exhibits higher modulation bandwidth (8.2GHz) than the ES (6.3GHz), the amplitude of ES oscillations is one order of magnitude greater than those of the GS level (Fig.3b black and red dashed line). This effect is observed, although the mean output power from each state is almost identical (fig.3a). The reason for this difference lies upon two main factors: the occupation probability of each state and the gain of each state. Specifically, the high occupation probability of the GS ($p_{e,GS} >$ 0.9) causes the gain to saturate and prohibits strong fluctuations of GS carrier due to the Pauli blocking effect [36]. On the other hand, ES occupation probability is lower ($p_{e,ES} > 0.6$). Therefore, fluctuations of higher amplitude are permitted in ES since the Pauli blocking effect is not so strong at this state while the higher gain of the ES state increases the amplitude of ES oscillations even more. These enhanced oscillations are anticipated to give rise to enhanced temporal dynamics, when ES state emission is included for computational tasks.

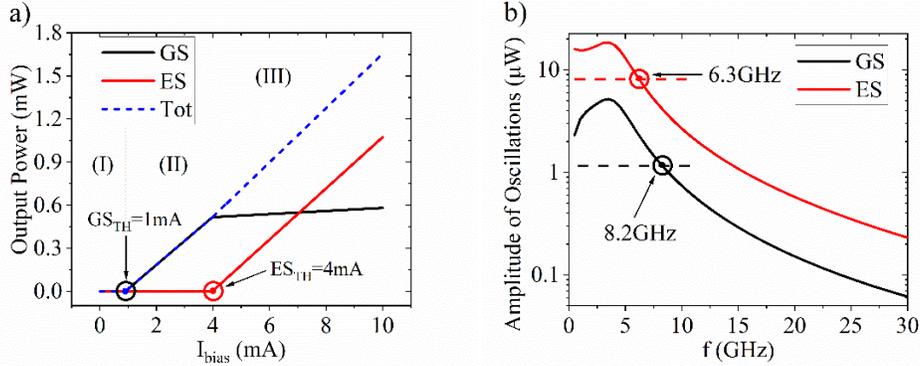

*Figure 3 (a) Relation between the applied current and the output power of the QD s-VCSEL from the GS (black), the ES (red) and both states (blue). A common bias is applied to both polarizations ($I^+ = I^-$). (b) The frequency response of the QD sVCSEL for a small signal analysis when the bias is set to 7mA (black for the GS and red for the ES).*

### 3.2 QD TDRC Simulation

The different bias conditions discussed above are now evaluated in the context of TDRC. The benchmark task assumed, is the equalization of a pulse-amplitude-modulated optical signal with four levels (PAM-4), transmitted at a rate of 25Gbaud/s at 1550nm, propagating through a 50km optical link using standard single-mode-fiber. The evaluation of the system is done through the counting of faulty retrieved symbols(bits), which determines the Bit Error Rate (BER) of the assigned task. Signal propagation effects are derived with a split-step Fourier method governed by Manakov equations [37], whereas as stated above noise effects at the detection stage are incorporated, similar to [11].

In order to solve this task, as we mention above, we focused our analysis on an external loop of *T=240ps* while the virtual nodes of the TDRC are separated by $T_n$=20ps. Consequently, the proposed QD TDRC scheme supports $N=T/T_n=10$ nodes and imposes an inherent speed penalty of $SP=T/T_S=6$. The injection strength is set to $r_{inj}=0.8$ while the feedback strength is kept to $r_f=0.01$ (see table I). In the QD TDRC we have access to four different optical fields, therefore, with regards to the masking procedure we can apply one and up to four different masks, at each optical field of the SL so as to improve the BER. To distinguish these cases at the graphs we inserted the following notation; $m$ is the number of masks applied to the emission states of the QD sVCSEL, while $p$ is the number of masks applied to its polarization states. Consequently, when $m=1$ and $p=2$, one common mask is applied to GS$^+$ and ES$^+$ and a second mask is applied to GS$^-$ and ES$^-$. The combinations of $m$ and $p$ which are used in this work are given in Table 2. The application of four masks $(m=2, p=2)$ comes with a hardware complexity since two polarization beam splitters and two frequency modulators must be added to the original layout (see Fig.2b). At the output layer of the TDRC, all four fields are always detected, sampled, digitized and send to the digital linear regression. Training and testing are based on a batch of 30000 PAM-4 symbols, 10000 for training and 20000 for testing. In order to validate training/inference, we repeat the whole procedure 10 times; each time we randomly choose 10000 symbols from the set of 30000, for training and use the rest for testing. In addition, the input of the Linear Regression algorithm comprises a batch of 21 consecutive symbols (taps), similar to [9].

Table 1 Combinations of m and p and the masking of the corresponding fields

| m (mask in each waveband) | p (mask in each polarization) | Number of different masks | Fields with the same masks | |
|---|---|---|---|---|
| 1 | 1 | 1 | i.GS$^+$, ES$^+$, GS$^-$ and ES$^-$ | |
| 1 | 2 | 2 | i.GS$^+$ and ES$^+$ | ii. GS$^-$ and ES$^-$ |
| 2 | 1 | 2 | i.GS$^+$ and GS$^-$ | ii.ES$^+$ and ES$^-$ |
| 2 | 2 | 4 | i.GS$^+$  ii.ES$^+$ | iii. GS$^-$  iv.ES$^-$ |

*3.3 Signal Equalization using single/multi-band QD nodes and single/multiple masks*

In this section, performance in the PAM-4 equalization is evaluated for different regimes of operation that correspond to different wavebands GS/ES versus a typical optimization factor, which in TDRCs is the frequency detuning (df) between ML and SL emission (fig.4). In addition, the impact of a varying number of pseudo-random masks at the available outputs is also investigated. Regime II, where only GS emission is achieved, can be considered the basis of the evaluation process and it is also equivalent to previous VCSEL based TDRCs [12-14]. It can be seen, that BER is capped at 3.5e-2 for a single mask (identical mask for both $GS^+$ and $GS^-$), whereas an improvement with a BER=1.6e-2, is achieved by using two different masks that allow the decorrelation of the two polarization states. In any case, performance is not compatible with HD-FED for this $T$ (fig.4a). On the other hand, it can be seen that if injection current is increased and the QD laser is biased at regime III, where ES lasing is present but weak, then even with a single mask *(m=1, p=1)* BER diminishes to 6.5e-3 (fig.4b). Furthermore, by increasing the number of masks to 2, *(m=2, p=1)* we further reduce BER to 3.7e-3, while by using four masks (fig.2b) performance is optimized to 1.4e-3, rendering for the first time QD TDRCs HD-FEC compatible. Two are the main reason behind this performance enhancement. On one hand, ES lasing, allow a second "parallel" signal in the TDRC's feedback loop, thus doubling the number of available RC nodes and boosting performance. The second reason, is the obvious improvement of modulation bandwidth with increasing injection current, due to the damping of relaxation oscillations. The same behavior can be observed when injection current, drives the laser to deep inside regime III. In this case ES emission is equally strong to GS, whereas modulation bandwidth is further improved. In this case, the combination of GS/ES emission with a single mask (m=1, p=1) results to BER=1.3e-3, which is similar to the 4-mask result at I=4.5mA, while if a 4-mask scheme is used a record level of BER=2.3e-4 is computed that is comparable to state-of-the-art multi-longitudinal QW fabry perot schemes [9]. It is worth mentioning, that in this regime, even a 2-mask scheme (m=2, p=1) can offer adequate performance (BER=4.9e-4) through a less demanding implementation.

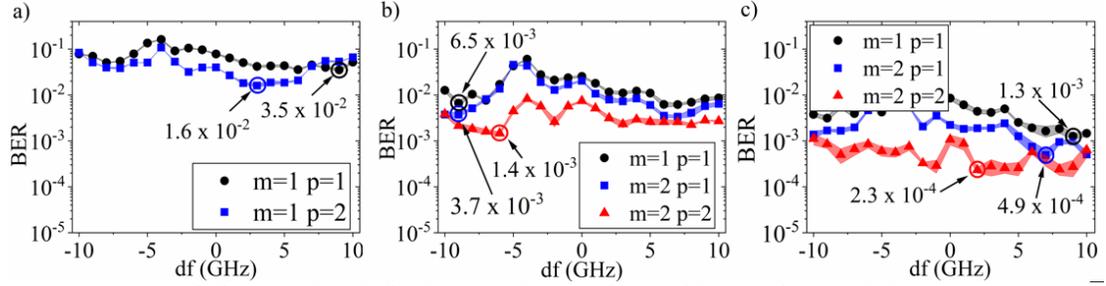

Figure 4 BER versus the ML-SL frequency detuning for $I_{bias}$=1.3mA – Regime I(left), $I_{bias}$=4.5mA – Regime II (center) and $I_{bias}$=7mA – Regime III (right). Black circles denote the use of a single mask at every state of the VCSEL while blue squares and red triangles stand for the use of two and four different masks at every optical field.

Aiming to further shed light on the explicit contribution of the ES emission on TDRC's performance, we made the following analysis. We biased the node at Ibias=7mA, deep inside regime III, where both GS and ES are present and modulation bandwidth is increased, but we used only the GS emission of the ML so as to feed the input symbols, removing optical injection from the ES $(k_{inj,ES}=0)$ (Fig.5a black squares) [35-36]. In Fig.5a the single waveband TDRC, although at the high injection current offers a BER=6.1e-3 which 10 fold higher compared to fig.4c (BER=4.9e-4). This performance deterioration can be solely attributed to the lack of dual-band injection. The same can be seen when information is loaded only at the ES band $(k_{inj,GS}=0)$ (fig.5a red-triangles). In this case, BER is reduced to 1.8e-3, lower compared to $k_{inj,ES}=0$ case, due to ES's enhanced temporal dynamics, but significantly higher compared to dual emission.

Lastly, we examined a case where aims at optimizing performance while preserving implementation simplicity. In this case, our motive is based on the observation that injection locking can enhance the bandwidth response of the SL. Therefore, we repeated the aforementioned procedure; meaning that input was injected to only one band (GS or ES) but now the non-modulated state was also injected to the SL $(k_{injGS} \neq 0, k_{injES} \neq 0)$ as CW. Under these conditions, a significant improvement can be observed in terms of BER compared to fig.5a, which emphasizes the beneficial role of the injection of the non-modulated field. In particular, again for the ES a BER of 6e-4 can be achieved with only two masks, thus generating equivalent performance with the deep $I_{bias}$=7mA, 4 mask case used in fig.4c.

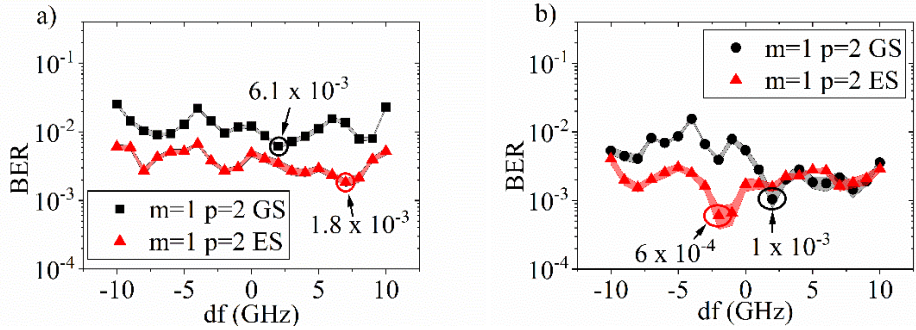

Figure 5 s-QDTDRC performance (a) when two polarized fields of the same quantum state GS or ES of the dual emitting QD spin VCSEL is phase modulated (black squares for GS and red triangles for the ES) (a) without injection of the non-modulated field ($k_{inj,GS}=0$, $k_{inj,ES}=0$) (b) with injection of the non-modulated field ($k_{injES} \neq 0$, $k_{injGS} \neq 0$).

### 3.4 Comparison with other similar works

In this section a comparison between the proposed QD s-VCSEL TDRC and other similar works based on quantum well spin VCSELs [38] and conventional Fabry-Perot lasers [9][39] is presented. The aim of this comparison is to highlight the merits of the proposed scheme. To begin with, in [38] the authors used a QW spin VCSEL as the basis of their TDRC. This point is of paramount importance due to the fact that authors in [38] have access to two "independent" modes (polarization states) that are exploited so as to boost the number of virtual nodes or solve two tasks in parallel. In our case, by using QD materials we unlock four modes; the two possible emitted wavebands (GS/ES) and at the same time the two polarization states of each waveband. Therefore, we have the capability to quadrable the virtual nodes or alternatively to solve 4 tasks in parallel with enhanced performance. The price to pay is the inclusion of four phase modulators and corresponding broadband filters to allow independent masking to each mode. It is worth stating that this is the first work according to our knowledge that all four degrees of freedom of a QD s-VCSEL are used as an RC concept. In the TDRC proposed in [9] our group exploited the dynamics between the quasi-independent longitudinal modes of a conventional Fabry Perot laser so as to tackle the channel equalization of a 25Gbaud PAM-4 signal transmitted through 50km. This work is directly compared to this one, due to the fact that the same benchmark test was assumed. The best BER value achieved was $3 \cdot 10^{-4}$ while the speed penalty of the network was minimized thanks to the small length of the external loop (240ps). Moreover, the authors managed to achieve HD-FEC compatibility in the parallel processing of two independent PAM-4 signals. In our case, identical results with [9] were obtained with the use of only four different modes which cuts the number of neurons at half. This fact clearly depicts the richness of QD s-VCSEL dynamics and may enable parallel processing in the future. Finally, an all-optical TDRC implementation was presented in [39] where the authors exploit the nonlinear dynamics of a conventional quantum well laser. In this case, the task was to equalize a 28Gbaud PAM-4 signal transmitted though 100km. Although the performance of this implementation is superior to previous schemes, since it relies on optical splitting into N optical taps, it requires high SNR and is more challenging if a real time training approach is adopted.

From the above comparison, it can be concluded that our approach compared to previous TDRC schemes draws its strength from the rich dynamics of the dual-band emitting QD s-VCSELs and the fact that four quasi-independent modes are utilized in parallel, thus quadrupling the number of virtual nodes that can be included in a single loop. Therefore, power consumption and complexity wise, the proposed scheme uses a single short delay loop and a low power VCSEL. On the other hand, polarization/broadband filters should be included so as to split the different modes, impacting complexity and four phase modulators that also slightly affect the overall power consumption. Finally, the proposed scheme can be further evolved exploiting new concepts in TDRC design. In particular, the replacement of the external loop by a multi-mode resonator can allow the time multiplexing of all virtual nodes through a twofold mechanism [40]. first the increase of the loop delay by enabling multiple propagation of light within the cavity and second the complex interaction between all modes. These two facts in combination with the QD s-VCSEL scheme can further enhance node dynamics, reduce feedback loop's size and boost performance.

## 4. Conclusion

A TDRC scheme is presented that uses a low power QD spin polarized VCSEL as a non-linear neural node. Multi band emission unlocked by the QD material is exploited for the first time so as to generate complex temporal dynamics and boost virtual node count, beyond the speed penalty limit of an integration ready TDRCs. Numerical simulations targeting the equalization of a 25Gbaud PAM-4 signal subject to dispersion after 50Km transmission revealed that by increasing QD spin polarized VCSEL's injection current and achieving dual emission, the number of virtual nodes increase, vastly surpassing typical single waveband QD performance. The paramount importance of including ES emission in the TDRC is confirmed, allowing for the first time, equivalent performance to state-of-the-art QW devices. Finally, the role of using multiple masks has been investigated, shedding light to the dimensionality expansion property of RCs, further boosting performance but at the cost of higher implementation complexity.

### Acknowledgments

This work has received funding and support from the EU H2020 NEOTERIC project (871330), whereas G. Sarantoglou has received funding from the Hellenic Foundation for Research and Innovation (HFRI) and the General Secretariat for Research and Technology (GSRT), under grant agreement No 2247 (NEBULA project).

**References**
1. Lukoševičius, M., & Jaeger, H. (2009). Reservoir computing approaches to recurrent neural network training. *Computer Science Review*, *3*(3), 127-149.
2. Jaeger, H. (2002). *Tutorial on training recurrent neural networks, covering BPPT, RTRL, EKF and the" echo state network" approach* (Vol. 5, No. 01, p. 2002). Bonn: GMD-Forschungszentrum Informationstechnik.
3. Jacobson, P. L., Shirao, M., Yu, K., Su, G. L., Wu, M. C. (2021). Hybrid Convolutional Optoelectronic Reservoir Computing for Image Recognition. *Journal of Lightwave Technology*. MNIST
4. Larger, L., Baylón-Fuentes, A., Martinenghi, R., Udaltsov, V. S., Chembo, Y. K., Jacquot, M. (2017). High-speed photonic reservoir computing using a time-delay-based architecture: Million words per second classification. *Physical Review X*, *7*(1), 011015.
5. Hou, Y. S., Xia, G. Q., Jayaprasath, E., Yue, D. Z., Yang, W. Y., Wu, Z. M. (2019). Prediction and classification performance of reservoir computing system using mutually delay-coupled semiconductor lasers. *Optics Communications*, *433*, 215-220.
6. Sozos, K., Bogris, A., Bienstman, P., Mesaritakis, C. (2021, September). Photonic Reservoir Computing based on Optical Filters in a Loop as a High Performance and Low-Power Consumption Equalizer for 100 Gbaud Direct Detection Systems. In *2021 European Conference on Optical Communication (ECOC)* (pp. 1-4). IEEE.
7. Tanaka, G., Yamane, T., Héroux, J. B., Nakane, R., Kanazawa, N., Takeda, S., Numata, H., Nakano, D., Hirose, A. (2019). Recent advances in physical reservoir computing: A review. *Neural Networks*, *115*, 100-123.
8. Brunner, D., Soriano, M. C., Mirasso, C. R., Fischer, I., "Parallel photonic information processing at gigabyte per second data rates using transient states," Nature Commun., vol. 4, 2013, Art. no. 1364.
9. Bogris, A., Mesaritakis, C., Deligiannidis, S., Li, P. (2020). Fabry-Perot lasers as enablers for parallel reservoir computing. *IEEE Journal of Selected Topics in Quantum Electronics*, *27*(2), 1-7.
10. Paquot, Y., Duport, F., Smerieri, A., Dambre, J., Schrauwen, B., Haelterman, M., Massar, S. (2012). Optoelectronic reservoir computing. *Scientific reports*, *2*(1), 1-6.

11. Héeroux, J. B., Numata, H., Kanazawa, N., Nakano, D. (2018, July). Optoelectronic reservoir computing with VCSEL. In *2018 International Joint Conference on Neural Networks (IJCNN)* (pp. 1-6). IEEE.
12. Vatin, J., Rontani, D., Sciamanna, M. (2019). Experimental reservoir computing using VCSEL polarization dynamics. *Optics express*, 27(13), 18579-18584.
13. Tan, X., Hou, Y., Wu, Z., & Xia, G. (2019). Parallel information processing by a reservoir computing system based on a VCSEL subject to double optical feedback and optical injection. *Optics express*, 27(18), 26070-26079.
14. Guo, X. X., Xiang, S. Y., Zhang, Y. H., Lin, L., Wen, A. J., & Hao, Y. (2019). Polarization multiplexing reservoir computing based on a VCSEL with polarized optical feedback. *IEEE Journal of Selected Topics in Quantum Electronics*, 26(1), 1-9.
15. Nguimdo, R. M., Verschaffelt, G., Danckaert, J., Van der Sande, G. (2015). Simultaneous computation of two independent tasks using reservoir computing based on a single photonic nonlinear node with optical feedback. *IEEE transactions on neural networks and learning systems*, 26(12), 3301-3307.
16. Antonik, P., Hermans, M., Duport, F., Haelterman, M., & Massar, S. (2016, March). Towards pattern generation and chaotic series prediction with photonic reservoir computers. In *Real-time Measurements, Rogue Events, and Emerging Applications* (Vol. 9732, p. 97320B). International Society for Optics and Photonics.
17. Shang, C., Wan, Y., Selvidge, J., Hughes, E., Herrick, R., Mukherjee, K., Duan, J., Grillot, F., Chow, J. E., Bowers, J. E. (2021). Perspectives on advances in quantum dot lasers and integration with Si photonic integrated circuits. *ACS photonics*, 8(9), 2555-2566.
18. Jung, Y., Shim, J., Kwon, K., You, J. B., Choi, K., Yu, K. (2016). Hybrid integration of III-V semiconductor lasers on silicon waveguides using optofluidic microbubble manipulation. *Scientific reports*, 6(1), 1-7.
19. Zilkie, A. J., Meier, J., Mojahedi, M., Poole, P. J., Barrios, P., Poitras, D., Rotter, J. T., Yang, C., Stintz, A., Malloy, K. J., Smith, P. W. E., Aitchison, J. S. (2007). Carrier Dynamics of Quantum-Dot, Quantum-Dash, and Quantum-Well Semiconductor Optical Amplifiers Operating at 1.55μm. *IEEE Journal of Quantum Electronics*, 43(11), 982-991.
20. Erneux, T., Viktorov, E. A., Mandel, P. (2007). Time scales and relaxation dynamics in quantum-dot lasers. *Physical Review A*, 76(2), 023819.
21. Gioannini, M., Sevega, A., Montrosset, I. (2006). Simulations of differential gain and linewidth enhancement factor of quantum dot semiconductor lasers. Optical and Quantum electronics, 38(4), 381-394., ISO 690,
22. Dochhan, A., Eiselt, N., Griesser, H., Eiselt, M., Olmos, J. J. V., Monroy, I. T., & Elbers, J. P. (2016, September). Solutions for 400 Gbit/s inter data center WDM transmission. In *ECOC 2016; 42nd European Conference on Optical Communication* (pp. 1-3). VDE.
23. Qasaimeh, O. (2015). Novel closed-form solution for spin-polarization in quantum dot VCSEL. *Optics Communications*, 350, 83-89.
24. Qasaimeh, O. (2015). Effect of doping on the polarization characteristics of spin-injected quantum dot VCSEL. *Optical and Quantum Electronics*, 47(3), 465-476.
25. Qasaimeh, O. (2008). Novel closed-form model for multiple-state quantum-dot semiconductor optical amplifiers. IEEE journal of quantum electronics, 44(7), 652-657.
26. Gioannini, M. (2012). Ground-state power quenching in two-state lasing quantum dot lasers. *Journal of Applied Physics*, 111(4), 043108.
27. Georgiou, P., Tselios, C., Mourkioti, G., Skokos, C., Alexandropoulos, D. (2021). Effect of excited state lasing on the chaotic dynamics of spin QD-VCSELs. *Nonlinear Dynamics*, 106(4), 3637-3646.
28. Lee, J., Oszwałdowski, R., Gøthgen, C., Žutić, I. (2012). Mapping between quantum dot and quantum well lasers: From conventional to spin lasers. Physical Review B, 85(4), 045314.
29. Adams, M. J., & Alexandropoulos, D. (2012). Analysis of quantum-dot spin-VCSELs. *IEEE Photonics Journal*, 4(4), 1124-1132.
30. Sarantoglou, G., Skontranis, M., Mesaritakis, C. (2019). All optical integrate and fire neuromorphic node based on single section quantum dot laser. *IEEE Journal of Selected Topics in Quantum Electronics*, 26(5), 1-10.
31. Van Tartwijk, G. H. M., & Lenstra, D. (1995). Semiconductor lasers with optical injection and feedback. *Quantum and Semiclassical Optics: Journal of the European Optical Society Part B*, 7(2), 87.
32. Sugawara, M., Mukai, K., Nakata, Y., Ishikawa, H., Sakamoto, A. (2000). Effect of homogeneous broadening of optical gain on lasing spectra in self-assembled In x Ga 1− x A s/G a A s quantum dot lasers. *Physical Review B*, 61(11), 7595.
33. J. Tatebayashi, M. Ishida, N. Hatori, H. Ebe, H. Sudou, A. Kuramata, M. Sugawara, Y. Arakawa, "Lasing at 1.28 μm of InAs–GaAs quantum dots with AlGaAs cladding layer grown by metal–organic chemical vapor deposition," *IEEE J. Sel. Topics Quantum Electron.*, vol. 11, no. 5, pp. 1027–1034, Sep./Oct. 2005
34. Lüdge, K., & Schöll, E. (2011) Temperature dependent two-state lasing in quantum dot lasers. In *2011 Fifth Rio De La Plata Workshop on Laser Dynamics and Nonlinear Photonics* (pp. 1-6). IEEE.
35. Arsenijević, D., Schliwa, A., Schmeckebier, H., Stubenrauch, M., Spiegelberg, M., Bimberg, D., V. Mikhelashvili, M., Eisenstein, G. (2014). Comparison of dynamic properties of ground-and excited-state emission in p-doped InAs/GaAs quantum-dot lasers. *Applied Physics Letters*, 104(18), 181101.
36. Röhm, A., Lingnau, B., & Lüdge, K. (2015). Ground-state modulation-enhancement by two-state lasing inquantum-dot laser devices. *Applied Physics Letters*, 106(19), 191102.
37. D. Marcuse, C. R. Menyuk, and P. K. A. Wai, "Application of the Manakov-PMD equation to studies of signal propagation in optical fibers with randomly varying birefringence," J. Lightw. Technol., vol. 15, no. 9, pp. 1735–1746, Sep. 1997.

38. Yang, Y., Zhou, P., Mu, P., & Li, N. (2022). Time-delayed reservoir computing based on an optically pumped spin VCSEL for high-speed processing. *Nonlinear Dynamics*, 1-14.
39. Li, S., Pachnicke, S. (2019, November). Optical Equalization using Photonic Reservoir Computing with Optical Analog Signal Injection. In Asia Communications and Photonics Conference (pp. T4G-5). Optical Society of America.
40. S. Boshgazi, A. Jabbari, K. Mehrany, and M. Memarian, "Virtual reservoir computer using an optical resonator" Material Optics Express, Vol.12 No. 3, pp. 1140-1153, (2022)